\documentclass[a4paper,11pt]{article}
\pdfoutput=1 \usepackage{jheppub} \usepackage[T1]{fontenc} \usepackage{bm}
\usepackage{braket}

\usepackage[finalizecache=false, frozencache=true, cachedir=./]{minted}  

\usepackage[colorinlistoftodos]{todonotes}

\usepackage[utf8]{inputenc}
\DeclareUnicodeCharacter{0394}{\ensuremath{\Delta}}

\title{An analytic result for the $0 \to ggHHH$ amplitude}
\author[a]{John M. Campbell,}
\emailAdd{johnmc@fnal.gov}
\author[b]{Giuseppe De Laurentis,}
\emailAdd{giuseppe.delaurentis@ed.ac.uk}
\author[c]{R. Keith Ellis,}
\emailAdd{keith.ellis@durham.ac.uk}

\affiliation[a]{Fermilab, PO Box 500, Batavia IL 60510-5011, USA}
\affiliation[b]{Higgs Centre for Theoretical Physics, University of Edinburgh, Edinburgh, EH9 3FD, UK}
\affiliation[c]{Institute for Particle Physics Phenomenology, Durham University, Durham, DH1 3LE, UK}

\preprint{FERMILAB-PUB-25-0453-T,\, IPPP/25/45} \abstract{
We present a fully analytic calculation of the leading-order
  one-loop amplitude for triple Higgs production via gluon fusion, $gg \to
  HHH$, retaining full dependence on the mass of the heavy quark
  circulating in the loop.  This amplitude provides a direct probe of
  the triple and quartic Higgs self-couplings, the measurement of which
  is a central goal of current and future colliders. The amplitude can be
  presented in compact form thanks to the use of analytic
  reconstruction techniques, based on finite-field and
  $p\kern0.2mm$-adic evaluations, partial fraction decompositions, and
  primary decompositions to identify common numerator factors.  Our
  results provide a compact and efficient representation of the matrix
  element for this process, enabling evaluations that are more than an
  order of magnitude faster than existing numerical alternatives.
}

\newcommand{\etal}{{\it et al.}}

\def\x{{\times}}

\def\three{{\bf 3}}
\def\four{{\bf 4}}
\def\five{{\bf 5}}

\def\beq{\begin{equation}}
\def\eeq{\end{equation}}
\def\beqn{\begin{eqnarray}}
\def\eeqn{\end{eqnarray}}
\def\nn{\nonumber}

\def\spa#1.#2{\left\langle#1\,#2\right\rangle}
\def\spb#1.#2{\left[#1\,#2\right]}
\def\spaa#1.#2.#3{\langle\mskip-1mu{#1} 
                  | #2 | {#3}\mskip-1mu\rangle}
\def\spbb#1.#2.#3{[\mskip-1mu{#1}
                  | #2 | {#3}\mskip-1mu]}
\def\spab#1.#2.#3{\langle\mskip-1mu{#1} 
                  | #2 | {#3}\mskip-1mu]}
\def\spba#1.#2.#3{[\mskip-1mu{#1} 
                  | #2 | {#3}\mskip-1mu\rangle}
\def\spaba#1.#2.#3.#4{\langle\mskip-1mu{#1} 
                  | #2 | #3 | {#4}\mskip-1mu\rangle}
\def\spbab#1.#2.#3.#4{[\mskip-1mu{#1} 
                  | #2 | #3 | {#4}\mskip-1mu]}
\def\spabab#1.#2.#3.#4.#5{\langle\mskip-1mu{#1}
                  | #2 | #3 | {#4}| {#5} \mskip-1mu]}
\def\spbaba#1.#2.#3.#4.#5{[\mskip-1mu{#1} 
                  | #2 | #3 | {#4}| {#5}\mskip-1mu\rangle}
\def\spababa#1.#2.#3.#4.#5.#6{\langle\mskip-1mu{#1}
                  | #2 | #3 | {#4}| {#5}| {#6} \mskip-1mu\rangle}
\def\spbabab#1.#2.#3.#4.#5.#6{[\mskip-1mu{#1} 
                  | #2 | #3 | {#4}| {#5}| {#6}\mskip-1mu]}
\def\tr#1.#2{\text{tr}(#1|#2)}

\def\slsh{\rlap{$\;\!\!\not$}}     \def\three{{\bf 3}}
\def\four{{\bf 4}}
\def\five{{\bf 5}}
\def\trfive{\text{tr}_5}
\def\mt{m}
\def\mh{M_H}

\begin{document} 
\maketitle
\flushbottom

\section{Introduction}
\label{sec:intro}
The measurement of the couplings of the Higgs boson remains one of the major
goals of particle physics. The standard model makes specific predictions about
the couplings of the Higgs boson to quarks and vector bosons. The self couplings
of the Higgs boson have a special role because of the information they give about
the vacuum and potentially about the behaviour of the vacuum at high temperature.
Within the standard model and with the measured Higgs boson mass
the transition between the broken and unbroken phases is a crossover.
The absence of a first order phase transition in the standard model means
that electroweak baryogenesis models must rely on physics beyond the standard model.
This argumentation emphasizes the importance of detailed information about the shape
of the Higgs potential at zero temperature, which in the current context relies on
measurements of the triple and quartic Higgs boson couplings.
We also refer the reader to the white paper~\cite{Abouabid:2024gms} which 
provides an overview of studies focussed on the determination of the quartic Higgs boson coupling.

A direct probe of the triple Higgs boson coupling is provided by the measurement of double
Higgs production, which at the LHC proceeds primarily through the one-loop process $gg \to HH$.  Similarly,
the process $gg \to HHH$ is sensitive to the quartic Higgs coupling, (as well as the triple Higgs coupling).
Clearly no unambiguous statement is possible about the quartic coupling using the
$gg \to HHH$ process without knowledge of the
top-Higgs boson coupling and the triple Higgs boson coupling.

The first calculations of the leading-order, one-loop, $gg \to HHH$ process were performed
20 years ago~\cite{Plehn:2005nk,Binoth:2006ym}.
These calculations emphasized the challenge of measuring the quartic Higgs coupling
even in $pp$~collisions at $\sqrt{s}=200$~TeV~\cite{Plehn:2005nk},
but noted that Beyond-the-Standard Models (BSM) could significantly impact the chances of observation.
Because of the potential for the $gg \to HHH$ process to constrain the quartic Higgs boson coupling
it is desirable to have a calculation of at least the NLO corrections~\cite{Huss:2025nlt}.  This
is particularly important since the cross section is proportional to $\alpha_s^2$ at leading order,
resulting in a large uncertainty in the theoretical prediction for the cross section
simply from the choice of renormalization scale.  However, since this process involves three massive
external states (of mass $\mh$) and an internal loop of massive quarks (of mass $\mt$), the two-loop virtual corrections that would
appear in the NLO calculation are unknown.

In the absence of exact two-loop results for the $gg\to HHH$ process, all calculations
that aim at NLO results (or indeed NNLO results) involve approximations.
Maltoni~\etal~perform an approximate NLO calculation~\cite{Maltoni:2014eza} based on reweighting results
computed in the Higgs-gluon effective field theory (HEFT, valid in the $\mt \to \infty$ limit) by
a factor based on matrix elements computed in the full theory (FT), $|M_{FT}|^2/|M_{HEFT}|^2$.  This
reweighting is performed separately for leading order-like and real radiation contributions. They
find that $K$-factors for the $gg\to HHH$ process decrease
with increasing energy, $1.61$ at 14~TeV falling to 1.35 at 100~TeV.
In refs.~~\cite{deFlorian:2016sit,deFlorian:2019app} de Florian~\etal~perform a NNLO calculation
in the heavy top limit, and augment it with top mass effects by reweighting
to produce an approximate NNLO result for the triple Higgs production cross section.
Applying these reweighting techniques to double Higgs boson production 
overestimates the exact
inclusive cross section by $32\%$ at $\sqrt{s}=100$~TeV and by $16\%$ at $\sqrt{s}=14$~TeV~\cite{Borowka:2016ypz}.
For differential distributions, the mass effects can be even larger. This illustrates the need for an exact NLO
triple Higgs boson production cross section.
While exact NLO results for this process remain beyond current
capabilities, recent progress in the computation of two-loop
five-point amplitudes -- both massless \cite{Abreu:2023bdp,
  Agarwal:2021vdh, Badger:2023mgf, Agarwal:2023suw,
  DeLaurentis:2023nss, DeLaurentis:2023izi} and with a single massive
leg \cite{Badger:2021nhg, Abreu:2021asb, Badger:2024sqv,
  Badger:2021ega, Hartanto:2022qhh, Badger:2022ncb, Badger:2024mir,
  DeLaurentis:2025dxw} -- suggests that exact analytic NLO matrix elements for
$gg\rightarrow HHH$ could eventually be obtained through numerical
integration-by-parts reduction over finite fields, followed by
analytic reconstruction of the master integral coefficients. This
prospect provides further motivation to obtain fully analytic one-loop
results as a foundational step. The analytic results presented here
are, in fact, obtained using a reconstruction method not too
dissimilar from the one that could eventually be employed at two
loops.

Direct measurements of Higgs production processes are not the only way to constrain the strength
of the cubic and quartic Higgs couplings.  Complementary indirect information on the triple Higgs boson
coupling can be obtained from measurements
of single Higgs production, due to the fact that the coupling enters in electroweak
corrections~\cite{McCullough:2013rea,Gorbahn:2016uoy,Degrassi:2016wml,Bizon:2016wgr,
Degrassi:2017ucl,DiVita:2017eyz,Kribs:2017znd,Maltoni:2017ims,DiVita:2017vrr,
Maltoni:2018ttu,Gorbahn:2019lwq}.  The quartic coupling can be similarly constrained through both
single and double Higgs
production~\cite{Maltoni:2018ttu,Bizon:2018syu,Liu:2018peg,Borowka:2018pxx,
Chiesa:2020awd,Gonzalez-Lopez:2020lpd,Stylianou:2023xit,Papaefstathiou:2023uum,
Heinrich:2024dnz,Dong:2025lkm,Haisch:2025pql}.

Current constraints from the LHC make use of the 
kappa framework~\cite{LHCHiggsCrossSectionWorkingGroup:2013rie,LHCHiggsCrossSectionWorkingGroup:2016ypw}
to allow for deviations from the Standard Model.  
In this approach the Higgs potential is written as,
\begin{equation}
V(H) = \frac{1}{2} \mh^2 H^2 + \kappa_3 \lambda v H^3 + \kappa_4 \frac{\lambda}{4}  H^4 
\end{equation}
where $\lambda = \mh^2/(2v^2)$ and $1/v^2 = \sqrt2 G_F$.  
The triple- and quartic-Higgs couplings are then
allowed to depart from their Standard Model values, $\kappa_3 = \kappa_4 = 1$.
The ATLAS collaboration has already been able to place limits on $\kappa_3$ from measurements of single
and double Higgs production~\cite{ATLAS:2022jtk}, $-0.4 < \kappa_3 < 6.3$,
(see also refs.~\cite{ATLAS:2024xcs,CMS:2024awa} for current results
and ref.~\cite{CMS:2025hfp} for projections for HL-LHC).
Aside from experimental constraints, the values of $\kappa_3$ and $\kappa_4$
can also be bounded by the requirement of perturbative
unitarity~\cite{DiLuzio:2017tfn,Stylianou:2023xit,Papaefstathiou:2023uum}.
The tree-level partial wave analysis in Ref.~\cite{Stylianou:2023xit} suggests
that $|\kappa_3| \lesssim 5$--$10$ and $|\kappa_4| \lesssim 60$. 

In this paper we will provide simple expressions for the leading-order 1-loop process,
$gg \to HHH$.   This will provide amplitudes with improved speed and stability that can be
used in an eventual NLO calculation in the full theory.  It will
also help to develop the necessary techniques and analytic understanding for a future computation of
the radiative corrections.  Since this process contains many massive particles, both as external
states and in loop propagators, it also provides a further milestone for the analytic reconstruction
program~\cite{Laurentis:2019bjh,DeLaurentis:2022otd,DeLaurentis:2023qhd,Campbell:2024tqg,Campbell:2025ftx}.
In terms of using this method for a $2 \to 3$ process with massive particles, this is
an ideal test case because the presence of so many scalar particles serves to limit the complexity of
the amplitude.

The outline of the paper is as follows.  In section~\ref{sec:overview} we
provide an overview of the calculation and present results for the
relatively simple sub-amplitudes involving triple and quartic Higgs couplings. 
Section~\ref{sec:pentagon} presents the calculation of the central result of
this paper, the contribution from genuine pentagon diagrams that do not contain
multi-Higgs couplings.  We present a cross-check of our results in
section~\ref{sec:results} before concluding in section~\ref{sec:conc}.
Appendix~\ref{Integrals} contains essential loop integral definitions for our
calculation.

\section{Overview}
\label{sec:overview}

We consider the process,
\begin{equation}
0 \rightarrow g(p_1) + g(p_2) + H(p_3) +  H(p_4) +  H(p_5) \,,
\end{equation}
with all the Higgs bosons on-shell, $p_3^2 = p_4^2 = p_5^2 = \mh^2$.

The diagrams contributing to triple Higgs production can be separated into three QCD gauge-invariant classes defined by
the number of propagators in the loop or, equivalently, the appearance (or not) of multi-Higgs interaction vertices.
The result for the full amplitude is given by,
\begin{eqnarray}
A_{\rm tot} = \delta^{AB} \frac{g_s^2}{16\pi^2} \, \frac{\mt^4}{v^3} \left(
A_3 + A_4 + A_5 \right)\, ,
\end{eqnarray}
where we have extracted overall color, coupling and loop factors, including
the mass of the quark circulating in the loop, $m$.  The sub-amplitudes $A_3$, $A_4$ and $A_5$ originate,
respectively, from triangle, box and pentagon diagrams as we will describe further below.

Each amplitude is first computed using the standard technique of
Passarino-Veltman reduction~\cite{Passarino:1978jh}. It is
cross-checked against an in-house semi-automated numerical unitarity
code (see~\cite{Ellis:2011cr} and references therein). The amplitude
is decomposed into a sum over basis integrals whose coefficients are
isolated and then simplified using analytic reconstruction through
finite-field and $p$-adic
evaluations~\cite{Laurentis:2019bjh,DeLaurentis:2022otd,DeLaurentis:2023qhd,Campbell:2024tqg,Campbell:2025ftx}. Univariate
$p\kern0.2mm$-adic slices at large ($\propto p^{-1}$) and small
($\propto p$) values of $m$, or in a finite field at generic values
of $m$, are used to obtain the least common denominators. Further
$p\kern0.2mm$-adic evaluations near codimension two surfaces are used
to determine both allowed partial fraction decompositions and possible
numerator factors, before fitting the ans\"atze. This simplification
step is the crucial ingredient that allows us to present the
amplitudes in analytic form in this paper.

We will present our amplitudes in terms of spinor products defined as,
\begin{gather}
\label{Spinor_products1}
\langle i\, j \rangle = \lambda_i^\alpha \lambda_{j,\alpha}
= \begin{pmatrix} \lambda_i^\alpha & 0 \end{pmatrix}
\begin{pmatrix} \lambda_{j,\alpha} \\ 0 \end{pmatrix}
= \bar{u}_-(p_i) u_+(p_j), \\
[ i\, j ] = \tilde\lambda_{i,\dot\alpha} \tilde\lambda_j^{\dot\alpha}
= \begin{pmatrix} 0 & \tilde\lambda_{i,\dot\alpha} \end{pmatrix}
\begin{pmatrix} 0 \\ \tilde\lambda_j^{\dot\alpha} \end{pmatrix}
= \bar{u}_+(p_i) u_-(p_j), \\
\langle i\, j \rangle [j\, i] = 2 p_i \cdot p_j,
\end{gather}
for lightlike momenta $p_i$ and $p_j$, as well as longer spinor strings containing non-lightlike momenta,
\begin{eqnarray}
\spab i.{{\bf k}}.j &=& \bar{u}_-(p_i) \, \slsh{p_k} \, u_-(p_j)\,, \\
\spaba i.{{\bf k}}.{{\bf l}}.j &=& \bar{u}_-(p_i) \, \slsh{p_k} \, \slsh{p_l} \, u_+(p_j)\,, \\
\spbab i.{{\bf k}}.{{\bf l}}.j &=& \bar{u}_+(p_i) \, \slsh{p_k} \, \slsh{p_l} \,  u_-(p_j)\,, \\
\spabab i.{{\bf k}}.{{\bf l}}.{{\bf m}}.j &=& \bar{u}_-(p_i) \, \slsh{p_k} \, \slsh{p_l} \, \slsh{p_m} \,  u_-(p_j)\,,
\end{eqnarray}
where
\begin{eqnarray}
\slsh{p_k} = \begin{pmatrix} 0 & k_{\alpha\dot\alpha} \\ k^{\dot\alpha\alpha} & 0\end{pmatrix} \, .
\end{eqnarray}
To remind the reader of the non-lightlike nature of the vectors in these spinor sandwiches we have
written them in boldface.
We will also use the standard notation for the kinematic invariants,
\begin{equation}
s_{ij} = (p_i+p_j)^2 \, ,
s_{ijk} = (p_i+p_j+p_k)^2 \,.
\end{equation}
The amplitudes also depend on the Gram determinants,
\begin{equation}
\label{eq:delta12x3x4}
\Delta_{12\x3\x4} = p_3.p_4(p_3.p_{12}p_4.p_{12}-s_{12} p_3.p_4)
 -\mh^2\left((p_3.p_{12})^2 + (p_4.p_{12})^2-\mh^2s_{12}\right) \,,
\end{equation}
and,
\begin{equation}
\label{eq:delta12x4}
\Delta_{12\x4} = (p_{12}.p_4)^2-s_{12}\mh^2 \,,
\end{equation}
as well as the quantity,
\begin{equation}
\label{eq:trfivedef}
\trfive = \text{Tr} \{ \slsh{p_1}\slsh{p_2}\slsh{p_3}\slsh{p_4}\gamma_5 \}
 = \spabab 1.\four.\three.2.1 - \spabab 1.2.\three.\four.1 \,.
\end{equation}

The analytic reconstruction is performed working in the covariant
polynomial quotient ring defined as,
\begin{equation}
  \mathbb{F}\big[ \lambda_{1, \alpha}, \; \tilde\lambda_{1, \dot\alpha}, \; \lambda_{2, \alpha},
   \; \tilde\lambda_{2, \dot\alpha}, \; p_{3, \alpha\dot\alpha}, \; p_{4, \alpha\dot\alpha},
    \; p_{5, \alpha\dot\alpha} \big] \,,
\end{equation}
modulo the equivalence relations imposed by,
\begin{equation}
  \big\langle p_3^2 - p_4^2, \; p_4^2 - p_5^2 \big\rangle \,,
\end{equation}
which enforces the on-shell relations $p_3^2=p_4^2=p_5^2=M_H^2$. This setup
is analogous to the ``scalar-tops'' construction described in
ref.~\cite[eqs.~(2.24),(2.25)]{Campbell:2025ftx}, but includes an extra
constraint on the masses. The Gr\"obner basis required to build
ans\"atze for the numerator polynomials is derived by eliminating
covariant variables in favour of invariant ones, like in
ref.~\cite[eqs.~(2.26)--(2.28)]{Campbell:2025ftx}. The resulting Gr\"obner
basis is slightly larger than in that previous computation.

We supplement previously computed primary decomposition with two new
ones involving spinor strings with three massive momenta sandwiched
between massless spinors. The ideal,
\begin{equation}
  \big\langle \, \langle 1|\five|\four|\three|2], \; \langle 2|\three|\four|\five|1] \,  \big\rangle \,,
\end{equation}
has two primary components that are also prime,
\begin{equation}\label{eq:primdec1}
  \big\langle \, \langle 1|\five|\four|\three|2], \; \langle 2|\three|\four|\five|1], \; \trfive \,
      \big\rangle \;\, \text{and} \;\, 
  \big\langle \, \langle 1|\five|\four|\three|2], \; \langle 2|\three|\four|\five|1], \; s_{15}, \; 
      s_{23} \, \big\rangle \, .
\end{equation}
The ideal,
\begin{equation}
  \big\langle \langle 1|\five|\four|\three|2] , \, \Delta_{12\x3\x4} \big\rangle \,,
\end{equation}
has at least four components, for three of which we have identified a
simple form,
\begin{equation}\label{eq:primdec2}
  \big\langle \, M_H, \; \five_{\alpha\dot\alpha}\four^{\dot\alpha\beta} \,
  \big\rangle \, , \;
\big\langle \, M_H, \; \four^{\dot\alpha\alpha}\three_{\alpha\dot\beta} \,
  \big\rangle \;\, \text{and} \;\, 
\big\langle \, \langle 1 | \three | 2], \; \langle 1 | \four | 2], \; \langle 1 | \three
      | \four | 1 \rangle, [2 | \three | \four | 2] \, \big\rangle \, \, .
\end{equation}
The latter ideal coincides with the one appearing in the decomposition
of $\big\langle \langle 1|\three|2] , \, \Delta_{12\x3\x4} \big\rangle$,
  see ref.~\cite[eq.~(2.48)]{Campbell:2025ftx}. The remaining component,
  which we conjecture to be primary, appears to be significantly more
  complex, and we are currently unable to provide a compact set of
  generators for it --- even though it can be computed via ideal
  quotients. In the ancillary files we provide a script,
  \texttt{test\_primary\_decompositions.py}, to check
  eq.~\eqref{eq:primdec1} and eq.~\eqref{eq:primdec2}.

\subsection{Triangles: triple- and quartic-Higgs couplings}

\begin{figure}
\begin{center}
\includegraphics[width=0.2\textwidth,angle=270]{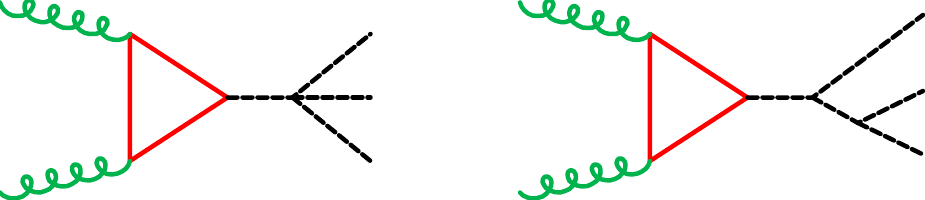}
\caption{Core triangle diagrams}
\label{triangles}
\end{center}
\end{figure}

The basic triangle diagrams which contribute to this amplitude are shown in Fig.~\ref{triangles}.
The diagram with a single quartic Higgs interaction occurs in 2 permutations and
the diagram with the two triple-Higgs couplings occurs in 6 permutations, giving 8 diagrams in all.

For opposite gluon helicities the triangle contribution to the amplitude
vanishes.  For equal helicities we must compute the coefficient of the triangle scalar integral $C_0(p_1,p_2,\mt)$
and a rational term.
We have,
\begin{eqnarray}
A_3^{++} &=& 
\frac{\spb1.2}{\spa1.2} \, \frac{6\mh^2}{\mt^2(s_{12}-\mh^2)} 
 \Bigl[(4\mt^2-s_{12}) C_0(p_1,p_2; \mt)+2\Bigr] \nn \\ 
 && \quad \times
 \left(\kappa_4+ \frac{3\kappa_3^2 \mh^2}{s_{34}-\mh^2}
  + \frac{3\kappa_3^2 \mh^2}{s_{35}-\mh^2} + \frac{3\kappa_3^2 \mh^2}{s_{45}-\mh^2} \right) \, ,
\\
A_3^{-+} &=& 0 \, .
\end{eqnarray}
The full definitions for the scalar triangle, box and pentagon integrals ($C_0,D_0$ and $E_0$) are given in Appendix~\ref{Integrals}.
The amplitude $A_3^{--}$ is obtained by replacing
$\spb1.2/\spa1.2$ with $\spa1.2/\spb1.2$ in the equation above.
 \subsection{Boxes: single triple-Higgs coupling}

\begin{figure}
\begin{center}
\includegraphics[width=0.2\textwidth,angle=270]{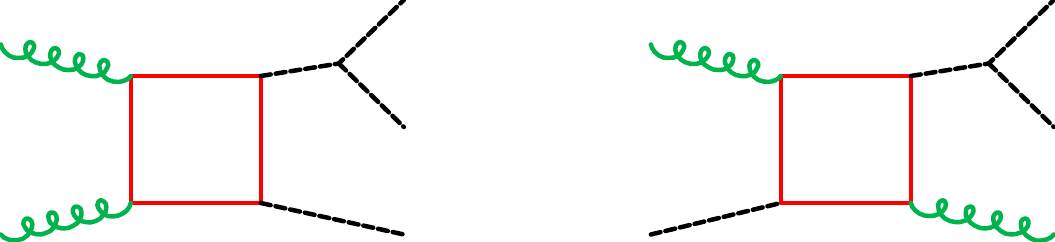}
\caption{Core box diagrams}
\label{box}
\end{center}
\end{figure}

The basic box diagrams that contribute to the amplitude are shown in Fig.~\ref{box}.
The diagram with the two adjacent gluons occurs with 12 permutations, whereas
the diagram with opposite gluons occurs	with only 6 permutations due to symmetry,
making 18 diagrams in all.

The full box amplitude can be expressed as a sum of scalar box integrals and scalar triangle integrals.
There is no contribution from scalar bubble integrals.
After permuting and summing we find that we can generate the full
amplitude by considering only 2 scalar box integrals:
\begin{itemize}
\item $D_0(p_2,p_1,p_3;\mt)$ $\times 6$ permutations,
\item $D_0(p_1,p_3,p_2;\mt)$ $\times 3$ permutations,
\end{itemize}
and only 4 scalar triangle integrals:
\begin{itemize}
\item $C_0(p_1,p_3;\mt)$ $\times 6$ permutations,
\item $C_0(p_1,p_{23};\mt)$ $\times 6$ permutations,
\item $C_0(p_3,p_{12};\mt)$ $\times 3$ permutations,
\item $C_0(p_1,p_2;\mt)$.
\end{itemize}
The coefficient of the triangle $C_0(p_3,p_{12};\mt)$ vanishes for the case of equal gluon helicities.

The coefficients for all of these integrals are relatively simple and one can
write the full subamplitude in a compact form. The result for equal gluon helicities is,
\begin{eqnarray}
\label{eq:A3Hpp}
A_4^{++} &=& \frac{\spb1.2}{\spa1.2} \, \frac{6\mh^2}{(s_{45}-\mh^2)}
\, \kappa_3
\Biggl\{ 
(8\*\mt^2-s_{12}-\mh^2-s_{45}) \, \*D_0(p_2,p_1,p_3; \mt) \nn \\
&&
+\left[\frac{(s_{13}\*s_{23}-\mh^2\*s_{45}+2\*s_{12}\*\mt^2)\*(8\*\mt^2-\mh^2-s_{45})}
 {4\*s_{12}\*\mt^2}
-\frac{s_{12}}{2} \right]
 \*D_0(p_1,p_3,p_2; \mt) \nn \\
&&
+\frac{(8\*\mt^2-\mh^2-s_{45})}{s_{12}\*\mt^2}
 \, \Bigl[ p_1\cdot p_{23} \, \*C_0(p_1,p_{23}; \mt)
         - p_1\cdot p_3 \, \*C_0(p_1,p_3; \mt) \Bigr]
\nn \\
&&
+2 \*C_0(p_1,p_2; \mt)
+\frac{1}{\mt^2}
\Biggr\} + \bigl(\text{5 perms}\bigr) \, .
\end{eqnarray}
Note that there is also a rational contribution for equal gluon helicities.

The sum over permutations corresponds to all combinations of $1 \leftrightarrow 2$ and cyclic
permutations of $(3,4,5)$, i.e. the total corresponds to the permutation sum
$P_6$ defined in eq.~\eqref{eq:P6} below.
Since the rational term and the ones involving $D_0(p_1,p_3,p_2; \mt)$ and $C_0(p_1,p_2; \mt)$
are invariant under the $1 \leftrightarrow 2$ swap, these terms could equally well
be written as a sum over three permutations and doubled rather than summed over all six permutations.

For opposite helicity gluons we have,
\begin{eqnarray}
\label{eq:A3Hmp}
A_4^{-+} &=& \frac{\spab1.\three.2}{\spab2.\three.1} \, \frac{6\mh^2}{(s_{45}-\mh^2)}
\, \kappa_3 \Biggl\{ 
\frac{(8\*\mt^2+s_{12}-\mh^2-s_{45})}{2}
 \*D_0(p_1,p_3,p_2; \mt) \nn \\
&&
+\left[(8\*\mt^2+s_{12}-\mh^2-s_{45})
       +\frac{s_{12}\*s_{13}\*(8\*s_{13}\*\mt^2-s_{13}^2-\mh^2\*s_{45})}
       {2\*\mt^2\*(s_{13}\*s_{23}-\mh^2\*s_{45})}
 \right] \*D_0(p_2,p_1,p_3; \mt)
 \nn \\
&&
+\frac{(8\*s_{23}\*\mt^2-s_{23}^2-\mh^2\*s_{45})}
       {\mt^2\*(s_{13}\*s_{23}-\mh^2\*s_{45})}
 \Bigl[ p_1\cdot p_{23} \, \*C_0(p_1,p_{23}; \mt)
      - p_2\cdot p_3 \, \*C_0(p_2,p_3; \mt) \Bigr]
 \nn \\
&&
+\frac{(s_{13}^2+s_{23}^2-2\*\mh^2\*s_{45})\*(8\*\mt^2+s_{12}-\mh^2-s_{45})}
       {4\*\mt^2\*(s_{13}\*s_{23}-\mh^2\*s_{45})}
 \, \*C_0(p_3,p_{12}; \mt)
 \nn \\
&&
-\frac{s_{12}\*(8\*(s_{13}+s_{23})\*\mt^2-s_{13}^2-s_{23}^2-2\*\mh^2\*s_{45})}
       {4\*\mt^2\*(s_{13}\*s_{23}-\mh^2\*s_{45})}
 \, \*C_0(p_1,p_2; \mt)
\Biggr\} + \bigl(\text{5 perms}\bigr)\, .
\end{eqnarray}
Note that in this sum over permutations the $1 \leftrightarrow 2$ operation should not be applied to the 
overall phase factor, $\spab1.\three.2/\spab2.\three.1$, in order to preserve the helicity of the gluons.
Equivalently, the $1 \leftrightarrow 2$ operation should be accompanied by charge conjugation, swapping
angle and square spinor brackets.
To simplify the result we have also traded the basis triangle integral $C_0(p_1,p_3; \mt)$ for
$C_0(p_2,p_3; \mt)$.
Again the coefficients of $D_0(p_1,p_3,p_2; \mt)$, $C_0(p_1,p_2; \mt)$ and $C_0(p_3,p_{12}; \mt)$ could equally well
be written as a sum over three permutations and doubled.

Amplitudes for gluons with both helicities flipped can be straightforwardly
obtained from these by interchanging square and angle spinor brackets.
Explicitly, $A_4^{--}$ is given by eq.~\eqref{eq:A3Hpp} with the factor
$\spb1.2/\spa1.2$ replaced by $\spa1.2/\spb1.2$.  Similarly,
$A_4^{+-}$ is obtained from eq.~\eqref{eq:A3Hmp} by replacing the factor
$\spab1.\three.2/\spab2.\three.1$ by $\spab2.\three.1/\spab1.\three.2$.

 \subsection{Pentagons: no multi-Higgs couplings}

\begin{figure}
\begin{center}
\includegraphics[width=0.2\textwidth,angle=270]{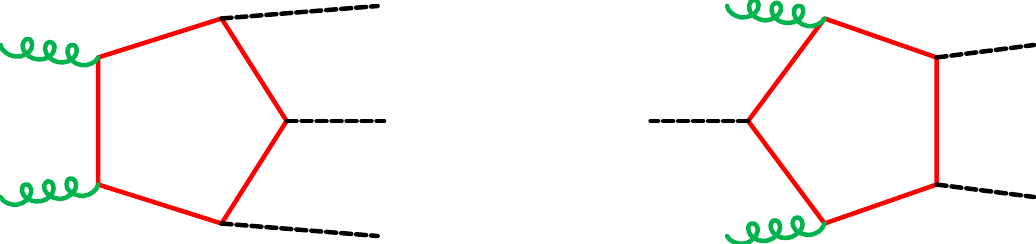}
\caption{Core pentagon diagrams}
\label{pentagon}
\end{center}
\end{figure}

There are 24 pentagon diagrams that contain no triple or quartic Higgs couplings.  These can be captured by computing two core
diagrams, with the gluons either adjacent or separated by a single Higgs boson, as shown in Fig~\ref{pentagon},
and thereafter accounting for a total of $2! \times 3! = 12$ permutations of the gluons and Higgs bosons.
There is no contribution from bubbles or a rational part.  After permuting and summing we find that we can generate the full
amplitude by considering only 5 boxes and 4 triangles.  
The full amplitude can be written as:
\begin{eqnarray}
A_5^{h_1h_2} &=& \sum_{P_{12}} d_{1\x23\x4}^{h_1h_2} \, D_0(p_1,p_{23},p_4; \mt)
  \nn \\
 &+& \sum_{P_6} \left[ d_{1\x2\x3}^{h_1h_2} \, D_0(p_1,p_2,p_3; \mt)
 + d_{1\x3\x24}^{h_1h_2} \, D_0(p_1,p_3,p_{24}; \mt) \right] \nn \\
 &+& \sum_{P_3} \left[ d_{1\x3\x2}^{h_1h_2} \, D_0(p_1,p_3,p_2; \mt)
 + d_{12\x3\x4}^{h_1h_2} \, D_0(p_{12},p_3,p_4; \mt) \right] \nn \\
 &+& \sum_{P_6} \left[ c_{1\x3}^{h_1h_2} \, C_0(p_1,p_3; \mt)
 + c_{1\x23}^{h_1h_2} \, C_0(p_1,p_{23}; \mt) \right] \nn \\
 &+& \sum_{P_3} c_{3\x12}^{h_1h_2} \, C_0(p_3,p_{12}; \mt)
 + c_{1\x2}^{h_1h_2} \, C_0(p_1,p_2; \mt)
\label{eq:amp}
\end{eqnarray}
Furthermore, the coefficients of the triangles $C_0(p_3,p_{12};\mt)$ and $C_0(p_1,p_2;\mt)$
vanish for the case of equal gluon helicities.
This sum involves three types of permutation over the momentum and helicity
labels:
\begin{eqnarray}
P_{12}: && (12345), (12453), (12534), (12354), (12543), (12435), \nn \\
\label{eq:P12}
        && (21345), (21453), (21534), (21354), (21543), (21435) \\
\label{eq:P6}
P_6: && (12345), (12453), (12534), (21345), (21453), (21534) \\
\label{eq:P3}
P_3: && (12345), (12453), (12534)
\end{eqnarray}
Note that for permutations that interchange $1$ and $2$ the helicity labels on
the coefficients also switch so that the amplitude $A^{-+}$ involves, for
instance, the  coefficients $c_{1\x3}^{-+}$ and $c_{2\x3}^{+-}$.  Coefficients
of integrals with gluons of reversed helicities are simply related by the
operation of complex conjugation or, equivalently, by interchanging
angle and square spinor brackets. For instance,
\begin{eqnarray}
c^{+-}_{1\x3} &=& \left[ c^{-+}_{1\x3} \right]^*
 = \left[ c^{-+}_{1\x3} \right]_{\spa{}.{} \leftrightarrow \spb{}.{}} \nn \\
c^{--}_{1\x3} &=& \left[ c^{++}_{1\x3} \right]^*
 = \left[ c^{++}_{1\x3} \right]_{\spa{}.{} \leftrightarrow \spb{}.{}}
\label{eq:cc}
\end{eqnarray}
We therefore only present results for two helicity combinations,
$++$ and $-+$, with the remainder inferred through eq.~\eqref{eq:cc}.
Since coefficients with permuted labels can be obtained by simply permuting
the momenta appearing in their expressions, this amplitude is fully specified
by the five box integrals and four triangle coefficients that are explicit
in eq.~\eqref{eq:amp}.

The calculation of the pentagon diagram contribution is the central result
of this paper and our results for the integral coefficients that appear
in eq.~\eqref{eq:amp} are presented in the following section.

\section{Pentagon contribution}
\label{sec:pentagon}

Since this amplitude contains pentagon diagrams the box coefficients that enter
the amplitude decomposition in eq.~\eqref{eq:amp} necessarily contain remnants of
the reduction from pentagon to box integrals.
We will first describe the procedure with which we handle this reduction
before presenting our results.

\subsection{Pentagon reduction}

To describe the pentagon integrals we follow the same approach
as in ref.~\cite{Campbell:2024tqg}.
The pentagon Gram determinant is defined by,
\begin{equation}
  \Delta(p_1,p_2,p_3,p_4)={\rm det}~G_{ij},\;\;\; G_{ij}=p_i \cdot p_j \,.
\end{equation}
It can be written in a simple form as,
\begin{equation}
  \Delta(p_1,p_2,p_3,p_4)= (\trfive)^2/16 \,,
\end{equation}
where $\trfive$ has been defined in eq.~\eqref{eq:trfivedef}.

Now define the Cayley matrix $S$ with elements,
\begin{equation}
 S_{ij} = m^2 - \frac{1}{2}(q_{i-1} - q_{j-1})^2 \,,
\end{equation}
where $q_i$ is the offset momentum in the pentagon integral.  An explicit expression for the
offset momenta is given in Appendix~\ref{Integrals}.
The scalar pentagon integral can be written as a sum of the 5 scalar box integrals
obtained by removing one denominator,
\begin{eqnarray}
  &&E_0(p_1,p_2,p_3,p_4;\mt)=
  c^{(1)} D_0(p_2,p_3,p_4;\mt)
  +c^{(2)} D_0(p_{12},p_3,p_4;\mt) \nn \\
  &+&c^{(3)} D_0(p_1,p_{23},p_4;\mt)
  +c^{(4)} D_0(p_1,p_2,p_{34};\mt)
  +c^{(5)} D_0(p_1,p_2,p_3;\mt)\, .
\end{eqnarray}
The reduction
coefficients of the pentagon into boxes are then given by~\cite{Bern:1993kr},
\begin{equation}
c^{(i)} = -\frac{1}{2} \sum_j S_{ij}^{-1} \,.
\label{eq:ci}
\end{equation}
Since this relation involves the inverse of $S$ these reduction coefficients
necessarily contain a denominator factor of the determinant of $S$.

\subsubsection{Adjacent gluons, $E_0(p_1,p_2,p_3,p_4,p_5;\mt)$}

In this case the relationship between the Cayley determinant and the Gram
determinant is,
\begin{equation} \label{eq:Cayley_and_Gram_adjacent}
16 \left|S^{1\x2\x3\x4\x5}\right| = 
 -s_{12} \, \spabab2.\three.\four.\five.1 \, \spabab1.\five.\four.\three.2 + \mt^2 (\trfive)^2 \,.
\end{equation}
The reduction coefficients can easily be obtained by explicit evaluation
of the expression in eq.~\eqref{eq:ci}.  However they can also be
written in a much more compact form using spinor notation.
The reduction coefficients are real quantities and can be written in terms of
traces without $\gamma_5$ matrices. They read,
\begin{eqnarray}
c^{(1)}_{1\x2\x3\x4} &=& -\frac{1}{2}\frac{\spaba2.\three.\four.2\spb1.2\spabab1.\five.\four.\three.2}{16 \left|S^{1\x2\x3\x4\x5}\right|}
 + \left\{\langle\rangle \leftrightarrow []\right\}
\, , \\
c^{(2)}_{1\x2\x3\x4} &=&
       \frac{1}{2}\frac{\spabab1.\five.\four.\three.2\*(\spb1.2\*\spaba2.\three.\four.2+\spa1.2\*\spbab1.\three.\four.1)}
       {16 \left|S^{1\x2\x3\x4\x5}\right|}
 + \left\{\langle\rangle \leftrightarrow []\right\}
      \nn \\ &&
       -\frac{1}{2} \, \frac{\trfive^2}{16 \left|S^{1\x2\x3\x4\x5}\right|}
\, , \\
c^{(3)}_{1\x2\x3\x4} &=& \left\{c^{(1)}_{1\x2\x3\x4}\right\}_{1 \leftrightarrow 2, 3 \leftrightarrow 5 }
\, , \\
c^{(4)}_{1\x2\x3\x4} &=&
       \frac{1}{2}\frac{s_{12}\*\spab2.\five.1\*\spabab1.\five.\four.\three.2}{16 \left|S^{1\x2\x3\x4\x5}\right|}
 + \left\{\langle\rangle \leftrightarrow []\right\}
\, ,\\
c^{(5)}_{1\x2\x3\x4} &=& \left\{c^{(4)}_{1\x2\x3\x4}\right\}_{1 \leftrightarrow 2, 3 \leftrightarrow 5 }\, .
\end{eqnarray}

\subsubsection{Non-adjacent gluons, $E_0(p_1,p_3,p_2,p_4,p_5;\mt)$}

The relationship between the Cayley determinant and the Gram
determinant is now,
\begin{equation}
16 \left|S^{1\x3\x2\x4\x5}\right| = 
\spab1.\three.2 \, \spab2.\three.1 \, \spaba2.\four.\five.1 \, \spbab2.\four.\five.1 + \mt^2 (\trfive)^2 \,.
\end{equation}
The corresponding reduction coefficients are,
\begin{eqnarray}
c^{(1)}_{1\x3\x2\x4} &=&
       \frac{1}{2}\frac{\spab2.\three.1\*\spbab2.\three.\four.2\*\spaba2.\four.\five.1}{16 \left|S^{1\x3\x2\x4\x5}\right|}
 + \left\{\langle\rangle \leftrightarrow []\right\}
\, , \\
c^{(2)}_{1\x3\x2\x4} &=&
       \frac{1}{2}\frac{\spab2.\three.1\*\spbab2.\four.\five.2\*\spaba1.\five.\four.2}{16 \left|S^{1\x3\x2\x4\x5}\right|}
 + \left\{\langle\rangle \leftrightarrow []\right\}
\, ,\\
c^{(3)}_{1\x3\x2\x4} &=& \left\{c^{(2)}_{1\x3\x2\x4}\right\}_{1 \leftrightarrow 2, 4 \leftrightarrow 5 }
\, ,\\
c^{(4)}_{1\x3\x2\x4} &=& \left\{c^{(1)}_{1\x3\x2\x4}\right\}_{1 \leftrightarrow 2, 4 \leftrightarrow 5 }
\, ,\\
c^{(5)}_{1\x3\x2\x4} &=&
       \frac{1}{2}\frac{\spb1.2\*\spab1.\three.2\*\spab2.\three.1\*\spaba2.\four.\five.1}{16 \left|S^{1\x3\x2\x4\x5}\right|}
 + \left\{\langle\rangle \leftrightarrow []\right\} \, .
\end{eqnarray}

\subsection{Effective pentagon and box coefficients}

Following the usual decomposition of the amplitude into scalar integrals leads to
both pentagon and box integrals.  The coefficients of both types of integrals contain
a single factor of the pentagon Gram determinant, leading to denominator
factors of $\trfive^2$ in both.  In order to facilitate cancellation of such factors between
the two integrals, after the reduction of the pentagons to boxes, we adopt the strategy of
ref.~\cite{Budge:2020oyl}.   For example, the coefficient of the pentagon integral with
adjacent gluons is rewritten as,
\begin{eqnarray}
\label{eq:effpent}
e_{1\x2\x3\x4} &=& 
\frac{16 \left|S^{1\x2\x3\x4\x5}\right| - \mt^2 (\trfive)^2 }
 {-s_{12} \, \spabab2.\three.\four.\five.1 \, \spabab1.\five.\four.\three.2} \, e_{1\x2\x3\x4} \\
&=& \left[ \frac{-16 \left|S^{1\x2\x3\x4\x5}\right|}
 {s_{12} \, \spabab2.\three.\four.\five.1 \, \spabab1.\five.\four.\three.2} \, e_{1\x2\x3\x4} \right]
 +\left[ \frac{ \mt^2 (\trfive)^2 }
 {s_{12} \, \spabab2.\three.\four.\five.1 \, \spabab1.\five.\four.\three.2} \, e_{1\x2\x3\x4} \right] \nn \,.
\end{eqnarray}
The first line is simply an insertion of a factor of unity resulting from the relation
between the Cayley and Gram determinants noted in eq.~\eqref{eq:Cayley_and_Gram_adjacent}.  The two terms that result are
interpreted in different ways.  The last term in square brackets is identified
as a genuine pentagon contribution;  this rescaled coefficient we call the ``effective
pentagon'' coefficient.  It contains no denominator factor of $\trfive^2$ since it is
explicitly cancelled in eq.~\eqref{eq:effpent}.  We absorb the first term in square
brackets into each of the box coefficients that results from this pentagon.  The explicit
numerator factor of $\left|S^{1\x2\x3\x4\x5}\right|$ cancels the corresponding factor
in the reduction coefficients, putting all the terms on the same footing and allowing
the remaining denominator factors of $\trfive^2$ to cancel.

Following this approach we find the results for the two effective
pentagon coefficients for the $-+$ helicity choice and for the cases of adjacent and non-adjacent gluons:
\begin{eqnarray}
e_{1\x2\x3\x4}^{-+} &=& \left\{
       -\frac{2\*\mt^2\*\trfive\*(\spaba1.\three.\four.1+\spab1.\three.2\*\spa1.2)}{\spa1.2\*\spabab2.\three.\four.\five.1}
       \right\}
       + \left\{ 1 \leftrightarrow 2, 3 \leftrightarrow 5, \spa{}.{} \leftrightarrow \spb{}.{} \right\}
       \nn \\ &&
       +\frac{4\*\mt^2\*\spabab1.\five.\four.\three.2\*(s_{12}-3\*\mh^2+8\*\mt^2)}{\spabab2.\three.\four.\five.1}
\end{eqnarray}

\begin{eqnarray}
e_{1\x3\x2\x4}^{-+} &=& \left\{
       \frac{2\*\mt^2\*\trfive\*\spaba1.\four.\five.1}{\spab2.\three.1\*\spaba2.\four.\five.1}
       \right\}
       + \left\{ 1 \leftrightarrow 2, 4 \leftrightarrow 5, \spa{}.{} \leftrightarrow \spb{}.{} \right\}
       \nn \\ &&
       +\frac{4\*\mt^2\*\spab1.\three.2\*(s_{12}-3\*\mh^2+8\*\mt^2)}{\spab2.\three.1}
\end{eqnarray}
Note that care must be taken in applying the symmetry operations in the first line of each of these
equations since they should also be applied to $\trfive$.  We have,
for instance, from eq.~\eqref{eq:trfivedef},
\begin{eqnarray}
\left\{ \trfive \equiv \trfive(p_1, p_2, p_3, p_4) \right\}_
{{1 \leftrightarrow 2, 3 \leftrightarrow 5, \spa{}.{} \leftrightarrow \spb{}.{}}}
&=&\text{Tr} \{ \slsh{p_2}\slsh{p_1}\slsh{p_5}\slsh{p_4}\gamma_5 \}_{\spa{}.{} \leftrightarrow \spb{}.{}} \nn \\
&=&\text{Tr} \{ \slsh{p_1}\slsh{p_2}\slsh{p_3}\slsh{p_4}\gamma_5 \}_{\spa{}.{} \leftrightarrow \spb{}.{}} \nn \\
&=&-\trfive(p_1, p_2, p_3, p_4) \,.
\end{eqnarray}

For the $++$ helicity choice, only a part of the pentagon coefficient has
a denominator factor of $\trfive^2$ so we only follow the above procedure
for this portion of the coefficient.  This leads to the following results for
the effective pentagon coefficients, for the cases of adjacent and non-adjacent gluons:
\begin{eqnarray}
e_{1\x2\x3\x4}^{++} &=&  \left\{
       \spb1.2\*(\tr3.4\*\spb1.2-\spbab1.\three.\four.2)
       +\frac{\mt^2\*\trfive\*(2\*\spbab1.\three.\four.1+\spb1.2\*\spab2.\four.1)}{\spa1.2\*\spabab2.\three.\four.\five.1}
       \right\}
       \nn \\ &&
       + \left\{ 1 \leftrightarrow 2, 3 \leftrightarrow 5 \right\}
       \nn \\ &&
       +\frac{4\*\mt^2\*\spb1.2\*(s_{12}-3\*\mh^2+8\*\mt^2)}{\spa1.2}
       +8\*\spb1.2^2\*\mt^2 \, ,
\end{eqnarray}

\begin{eqnarray}
e_{1\x3\x2\x4}^{++} &=&  \left\{
       \frac{2\*\mt^2\*\trfive\*(\spb1.2\*\spab2.\four.1-\spbab1.\three.\four.1)}{\spab2.\three.1\*\spaba2.\four.\five.1}
       \right\}
       + \left\{ 1 \leftrightarrow 2, 4 \leftrightarrow 5 \right\}
       \nn \\ &&
       +\frac{4\*\mt^2\*\spbab2.\four.\five.1\*(s_{12}-3\*\mh^2+8\*\mt^2)}{\spaba2.\four.\five.1}
       +2\*\spb1.2\*\spbab2.\four.\five.1+8\spb1.2^2\*\mt^2 \, .
\end{eqnarray}
Here we have introduced the notation,
\begin{equation}
\tr i.j = 2 p_i \cdot p_j \;.
\end{equation}
More generally we will use
\begin{equation}
\label{eq:trdef}
\text{tr} (a|b|\dots) = p_{a,\alpha\dot\alpha} p_{b}^{\dot\alpha\beta} \dots 
\end{equation}
where the number of arguments must be even in order to close the trace.

The box coefficients can then be written in terms of these effective pentagon
coefficients and a remainder.  Making use of the overall symmetry of the
coefficients under exchange of pairs of Higgs bosons, where possible, we finally
arrive at:
\begin{eqnarray}
\label{eq:effpent1}
d_{1\x2\x3} &=& \left\{ c_{1\x2\x3\x4}^{(5)} \,e_{1\x2\x3\x4}
 + \hat d_{1\x2\x3} \right\} + \left\{ 4 \leftrightarrow 5 \right\} \,, \\
\label{eq:effpent2}
d_{1\x3\x2} &=& \left\{ c_{1\x3\x2\x4}^{(5)} \,e_{1\x3\x2\x4}
 + \hat d_{1\x3\x2} \right\} + \left\{ 4 \leftrightarrow 5 \right\} \,, \\
\label{eq:effpent3}
d_{12\x3\x4} &=& \left\{ c_{1\x2\x3\x4}^{(2)} \,e_{1\x2\x3\x4}
 + \hat d_{12\x3\x4} \right\} + \left\{ 3 \leftrightarrow 5 \right\} \,, \\
\label{eq:effpent4}
d_{1\x3\x24} &=& \left\{ c_{1\x3\x2\x4}^{(4)} \,e_{1\x3\x2\x4}
 + \hat d_{1\x3\x24} \right\} + \left\{ 3 \leftrightarrow 5 \right\} \,, \\
\label{eq:effpent5}
d_{1\x23\x4} &=& 
 c_{1\x2\x3\x4}^{(3)} \,e_{1\x2\x3\x4} + c_{1\x3\x2\x4}^{(3)} \,e_{1\x3\x2\x4}
 + \hat d_{1\x23\x4} \,. 
\end{eqnarray}

\subsection{Box coefficient remainders}
The box remainders entering the decompositions in
eqs.~\eqref{eq:effpent1}--\eqref{eq:effpent5} are, for the opposite helicity case,
given by,
\begin{eqnarray}
\hat d_{1\x2\x3}^{-+} &=& 
       \frac{4\*s_{12}\*s_{23}^2}{\spab2.\three.1^2}
       +\frac{8\*\spab1.\three.2\*\mt^2}{\spab2.\three.1}
       -\frac{2\*\spab2.\three.2\*\spab1.(2+\three).1\*s_{23}\*(s_{12}+\mh^2)}{\spab2.\three.1\*\spabab2.\three.\four.\five.1}
       \nn \\ &&
       +\frac{8\*s_{12}\*s_{23}^2\*(\mh^2-2\*\mt^2)}{\spab2.\three.1\*\spabab2.\three.\four.\five.1}
       +\frac{s_{23}\*\spab1.\three.2\*(s_{124}-s_{12}+\mh^2)}{\spabab2.\three.\four.\five.1}
       \nn \\ &&
       -\frac{s_{23}\*\spab1.\four.2\*(s_{123}+s_{12}-\mh^2)}{\spabab2.\three.\four.\five.1} \, ,
\end{eqnarray}
\begin{eqnarray}
\hat d_{1\x3\x2}^{-+} &=& \spab1.\three.2 \left(
       \frac{8\*\mt^2}{\spab2.\three.1}
       +\frac{\spaba1.\four.\five.1}{\spaba2.\four.\five.1}
       +\frac{\spbab2.\four.\five.2}{\spbab2.\four.\five.1} \right)\, ,
\end{eqnarray}
\begin{eqnarray}
\hat d_{1\x3\x24}^{-+} &=&
       \frac{2\*\spab1.\three.1\*(s_{12}-3\*\mh^2+8\*\mt^2)\*(\spab1.\five.1\*\mh^2+\spab1.\three.1\*s_{15})}
        {\spa1.2\*\spab2.\three.1\*\spbab1.\three.\five.1}
       \nn \\ &&
       +\frac{(\spab1.\three.1\*\spab1.\four.2-s_{13}\*\spab1.\five.2)}{\spab2.\three.1}
       -\frac{2\*\spab1.\four.2\*(\spab1.\five.1\*\mh^2+\spab1.\three.1\*s_{15})}{\spa1.2\*\spbab1.\three.\five.1}
       \nn \\ &&
       +\frac{\spaba1.\three.\four.1\*\spab1.\three.1}{\spa1.2\*\spab2.\three.1}
       -\frac{\spab1.\three.2\*\spbab1.\three.\four.2\*\mh^2}{\spab2.\three.1\*\spbab2.\four.\five.1}
       -\frac{\spaba1.\three.\five.1\*\spaba1.\four.\five.1}{\spa1.2\*\spaba2.\four.\five.1}\, ,
\end{eqnarray}
\begin{eqnarray}
\hat d_{1\x23\x4}^{-+} &=& 
       \frac{16\*s_{23}\*\mt^2\*\spa1.2\*\spbab2.\four.\five.1}{\spab2.\three.1\*\spabab2.\three.\four.\five.1}
       +\frac{16\*s_{23}\*\mt^2\*\spb1.2\*\spaba2.\four.\five.1}{\spab2.\three.1\*\spabab2.\three.\four.\five.1}
       +\frac{16\*s_{23}\*\mt^2\*s_{12}\*(2\*\mh^2+\spab2.\four.2)}{\spab2.\three.1\*\spabab2.\three.\four.\five.1)}
       \nn \\ &&
       +\frac{16\*s_{23}\*\mt^2\*(\mh^4+\mh^2\*(-s_{12}-s_{13}+\spab1.\four.1)-\spab1.\three.1\*\tr3.4)}{\spab2.\three.1\*\spabab2.\three.\four.\five.1}
       \nn \\ &&
       -\frac{\spabab1.\five.\four.\three.1\*\spaba1.\four.\five.1}{\spab2.\three.1\*\spaba2.\four.\five.1}
       -\frac{\spb1.2\*\spab1.\three.2\*\mh^4}{\spab2.\three.1\*\spbab2.\four.\five.1}
       -\frac{\spbab1.\three.\four.2\*\spabab1.\five.\four.\three.2}{\spb1.2\*\spabab2.\three.\four.\five.1}
       -\frac{\spab1.\three.1\*\spaba1.\three.\four.1}{\spa1.2\*\spab2.\three.1}
       \nn \\ &&
       +\frac{((\spab1.\three.2\*(\mh^2-\spab2.\four.2-2\*s_{123})-\spab1.\four.2\*s_{23}-4\*\spab1.\four.2\*(s_{123}-s_{23}-\mh^2)))}{\spab2.\three.1}
       \nn \\ &&      
       -\frac{s_{123}\*(3\*\spab1.\three.2\*s_{15}+\spab1.\four.2\*s_{23}+4\*\spab1.\four.2\*(s_{15}+\mh^2)+\spab1.\three.1\*\spab1.\four.2)}{\spabab2.\three.\four.\five.1}
       \nn \\ &&
       +\frac{s_{23}\*(8\*\mh^2\*(\spab1.\four.2-\spab1.\three.2)+\spab1.\three.2\*(s_{13}+s_{23}+3\*s_{12})-3\*\spab1.\four.2\*s_{14}-2\*\spab1.\four.2\*s_{24})}{\spabab2.\three.\four.\five.1}
       \nn \\ &&
       +\frac{\mh^2\*\spab1.\four.2\*(4\*\mh^2+2\*\spab1.\three.1-4\*\spab1.\four.1)}{\spabab2.\three.\four.\five.1}
       -\frac{\spab1.\three.1\*\spab1.\four.2\*s_{14}}{\spabab2.\three.\four.\five.1}
       \nn \\ &&
       +\frac{2\*s_{23}\*(s_{12}-3\*\mh^2)\*s_{12}\*(-\mh^2+\spab1.\three.1+s_{123})}{\spab2.\three.1\*\spabab2.\three.\four.\five.1}
       -\frac{s_{23}\*(\spaba2.\three.\four.2\*\spbab2.\three.\four.2+4\*\Delta_{12\x3\x4})}{\spab2.\three.1\*\spabab2.\three.\four.\five.1}
       \nn \\ &&
       +\frac{2\*s_{23}\*(s_{12}-3\*\mh^2)\*(\spab1.\three.1^2-(\tr3.4+\spab1.\four.1)\*\spab1.\four.1)}{\spab2.\three.1\*\spabab2.\three.\four.\five.1} \, ,
\end{eqnarray}
\begin{eqnarray}
\hat d_{12\x3\x4}^{-+} &=&
      \frac{((\spaba1.\four.\five.1-\spaba1.\four.\three.1)\*\spabab1.\three.\four.\five.2\*(s_{12}-3\*\mh^2+8\*\mt^2))}{2\*\spa1.2\*\Delta_{12\x3\x4}}
       \nn \\ &&
       +(\spabab1.\three.\four.\five.1-\spabab2.\three.\four.\five.2)\*\Biggl[
        -\frac{(\spab1.\three.2\*\spaba1.\three.\five.1-2\*\spab1.\four.2\*\spaba1.\three.\four.1)}{4\*\spa1.2\*\Delta_{12\x3\x4}}
       \nn \\ &&
         \frac{\spabab1.\three.\four.\five.1\*\spaba1.\four.\five.1\*(2\*(s_{12}-3\*\mh^2+8\*\mt^2)+(s_{14}-2\*s_{15}-2\*s_{12}+2\*s_{34}))}
	 {4\*\spa1.2\*\spabab2.\three.\four.\five.1\*\Delta_{12\x3\x4}}
       \nn \\ &&
        -\frac{\spabab2.\three.\four.\five.2\*\spaba1.\three.\four.1\*(2\*(s_{12}-3\*\mh^2+8\*\mt^2)+(s_{14}-s_{15}-2\*s_{12}+2\*s_{34}-2\*\mh^2))}
	{4\*\spa1.2\*\spabab2.\three.\four.\five.1\*\Delta_{12\x3\x4}}
       \nn \\ &&
        -\frac{\spabab1.\three.\four.\five.2\*\spaba1.\three.\five.2\*(2\*(s_{12}-3\*\mh^2+8\*\mt^2)+(\mh^2-3\*s_{12}+2\*s_{14}-2\*s_{23}-s_{24}))}
	{4\*\spa1.2\*\spabab2.\three.\four.\five.1\*\Delta_{12\x3\x4}}
       \nn \\ &&
        +\frac{\spabab1.\three.\four.\five.2\*(3\*\spab1.\three.1^2+3\*\spab1.\four.1^2+2\*\tr3.4\*\spab2.\three.2+\spab2.\three.2^2)}{8\*\spabab2.\three.\four.\five.1\*\Delta_{12\x3\x4}}
       \nn \\ &&
        -\frac{\spaba1.\three.\four.1\*\mh^2\*(\mh^2\*\spab1.\four.1+(s_{34}-\mh^2)\*\spab2.\three.2)}{4\*\spa1.2\*\spabab2.\three.\four.\five.1\*\Delta_{12\x3\x4}}
        -\frac{\spab1.\four.2\*\spab1.\four.1\*\spabab1.\three.\four.\five.1}{2\*\spabab2.\three.\four.\five.1\*\Delta_{12\x3\x4}}
	\Biggr]
       \nn \\ &&
       + \left\{1 \leftrightarrow 2, 3 \leftrightarrow 5,
                \spa{}.{} \leftrightarrow \spb{}.{} \right\}\, .
\end{eqnarray}
Note that the expressions for the last two coefficients involve $\Delta_{12\x3\x4}$ which has been
introduced in eq.~\eqref{eq:delta12x3x4}.

For the equal helicity case they are,
\begin{eqnarray}
\hat d_{1\x2\x3}^{++} &=&       
       \frac{\spb1.2\*s_{23}\*(2\*\spbab2.\three.\four.2-\spab1.\four.2\*\spb1.2)}{2\*\spabab1.\five.\four.\three.2}
       -\frac{\spb1.2\*s_{23}\*(2\*\spbab1.\three.\four.1+\spab2.\four.1\*\spb1.2)}{2\*\spabab2.\three.\four.\five.1}
       \nn \\ &&
       -\frac{\spb1.2\*(s_{12}-8\*\mt^2)}{\spa1.2}\, ,
\end{eqnarray}
\begin{eqnarray}
\hat d_{1\x3\x2}^{++} &=&  \left\{
       -\frac{(\spab2.\three.2\*\spbab1.\three.\four.2)}{\spaba2.\four.\five.1}
       +\frac{(\spb1.2\*\spab1.\three.1\*(s_{14}-2\*\mh^2))}{\spaba2.\four.\five.1}
       -\frac{\spb1.2\*\spab1.\four.1\*s_{45}}{\spaba2.\four.\five.1} \right\}
       \nn \\ &&
       + \left\{1 \leftrightarrow 2, 4 \leftrightarrow 5 \right\}
       \nn \\ &&
       -\frac{4\*(\spab1.\three.2\*\spab2.\three.1+2\*s_{12}\*\mt^2)}{\spa1.2^2}
       -\frac{2\*(s_{12}-3\*\mh^2+8\*\mt^2)\*\spab1.\three.2\*\spab2.\three.1}{\spa1.2\*\spaba2.\four.\five.1}
       \nn \\ &&
       -\frac{2\*\spb1.2\*(2\*\spab1.\three.2\*\spab2.\three.1+s_{12}\*s_{45})}{\spaba2.\four.\five.1}\, ,
\end{eqnarray}
\begin{eqnarray}
\hat d_{1\x3\x24}^{++} &=&
       \frac{2\*(s_{12}-3\*\mh^2+8\*\mt^2)\*(\spab1.\three.1\*\mh^2+\spab1.\five.1\*s_{13})}{\spa1.2\*\spaba2.\four.\five.1}
       +\frac{\spbab1.\three.\four.1\*(s_{12}-s_{13}+\mh^2)}{\spa1.2\*\spab2.\three.1}
       \nn \\ &&
       -\frac{2\*\spab1.\three.2\*\spab1.\five.1\*s_{24}}{\spa1.2\*\spaba1.\three.\five.1}
       +\frac{(\spb1.2\*(2\*(s_{13}+s_{24})-(s_{12}+s_{34}))+4\*\spbab1.\three.\four.2)}{\spa1.2}
       \nn \\ &&
       +\frac{(4\*\Delta_{12\x3\x4}+\spaba2.\three.\four.2\*\spbab2.\three.\four.2)}{\spa1.2\*\spaba2.\four.\five.1}
       -\frac{\spab1.\five.2\*\spbab2.\four.\five.1\*(s_{13}-2\*\mh^2)}{\spab1.\three.2\*\spaba2.\four.\five.1}
       \nn \\ &&
       +\frac{\spb1.2\*(\mh^2\*(s_{134}-s_{13})+s_{13}\*(s_{23}+s_{15})-s_{24}\*(s_{12}+s_{14}-s_{34})-\mh^4)}{\spaba2.\four.\five.1}
       \nn \\ &&
       +\frac{\spbab1.\three.\four.2\*(s_{23}-s_{124})}{\spaba2.\four.\five.1}\, ,
\end{eqnarray}
\begin{eqnarray}
\hat d_{1\x23\x4}^{++} &=&
       \frac{\spbab1.\four.\five.1\*(s_{13}-s_{12}-\mh^2)}{\spa1.2\*\spab2.\three.1}
       +\frac{\spbab1.\four.\five.1\*(s_{12}\*(s_{23}-s_{15})-2\*(\mh^2\*s_{23}-s_{15}\*s_{45}))}{2\*\spa1.2\*\spabab2.\three.\four.\five.1}
       \nn \\ &&
       -\frac{(\spbab2.\four.\five.2\*s_{15}\*s_{45}+\spbab2.\three.\five.2\*s_{23}\*\mh^2)}{2\*\spa1.2\*\spabab1.\five.\four.\three.2}
       -\frac{\spbab2.\four.\five.1\*\spab1.\four.2\*(s_{45}-2\*s_{23})}{\spab1.\three.2\*\spaba2.\four.\five.1}
       \nn \\ &&
       +\frac{(\mh^2\*s_{23}-s_{15}\*s_{45})\*(s_{35}-s_{12}-6\*\mh^2+16\*\mt^2)}{\spa1.2\*\spaba2.\four.\five.1}
       \nn \\ &&
       -\frac{(s_{14}-\mh^2)\*\mh^4}{\spa1.2\*\spaba2.\four.\five.1}
       -\frac{(\spbab1.\three.\four.2+\spb1.2\*(s_{14}+s_{24}))}{2\*\spa1.2}
       \nn \\ &&
       -\frac{(\spbab2.\four.\five.1\*(s_{34}+s_{24}-s_{14}-2\*\mh^2)-\mh^2\*(\spbab2.\three.\five.1+\spb1.2\*\mh^2))}{\spaba2.\four.\five.1}\, ,
\end{eqnarray}
\begin{eqnarray}
\hat d_{12\x3\x4}^{++} &=& 
       -\frac{(\spbab2.\three.\four.2+\spbab2.\four.\five.2)\*\text{tr}(5|4|3|1-2)}{2\*\spa1.2\*\spabab1.\five.\four.\three.2}
       \nn \\ &&
       -\frac{(s_{34}-2\*\mh^2)\*(s_{35}+\mh^2-2\*s_{24})}{\spa1.2^2}
       -\frac{2\*((s_{13}-s_{23})\*\mh^2+\spaba1.\three.\four.2\*\spb1.2)}{\spa1.2^2}
       \nn \\ &&
       -\frac{(s_{45}-s_{34})\*(s_{13}-s_{23})}{16\*\spa1.2^2\*\Delta_{12\x3\x4}}\*\Bigl[(s_{45}-s_{34})\*(s_{13}+s_{23})\*(\tr{1+2}.4+4\*s_{34}-8\*\mt^2)
       \nn \\ && \quad
          +4\*(s_{34}-\mh^2)\*((s_{45}-s_{34})\*\tr{1+2}.4+s_{34}\*(s_{13}+s_{23}-s_{34}+2\*\mh^2)-8\*s_{123}\*\mt^2)\Bigr]
       \nn \\ &&
       +\frac{(\Delta_{12\x4}\*(s_{13}-s_{23})\*(s_{13}+s_{23})\*(\tr{1+2}.4-8\*\mt^2))}{4\*\spa1.2^2\*\Delta_{12\x3\x4}}
       \nn \\ &&
       + \left\{1 \leftrightarrow 2, 3 \leftrightarrow 5 \right\}\, .
\end{eqnarray}
In the last coefficient we have introduced, c.f. eq.~\eqref{eq:trdef},
\begin{equation}
\text{tr}(5|4|3|1-2) = \spabab1.\five.\four.\three.1-\spabab2.\five.\four.\three.2 \,.
\end{equation}
and $\Delta_{12\x4}$ has been defined in eq.~\eqref{eq:delta12x4}.

\subsection{Triangle coefficients}
The triangle coefficients appearing in eq.~\eqref{eq:amp}
are very simple.
For gluons with opposite helicities we have,
\begin{eqnarray}
c^{-+}_{1\times3} &=& -\frac{8\*\spab1.\three.1\*s_{13}}{\spab2.\three.1^2} \,, \\
c^{-+}_{1\times23} &=& \frac{8\*\spab1.(2+\three).1\*s_{23}}{\spab2.\three.1^2} \,, \\
c^{-+}_{3\times12} &=& \frac{8\*(\spab1.\three.1^2+\spab2.\three.2^2-2\mh^2 s_{12})}{\spab2.\three.1^2} \,, \\
c^{-+}_{1\times2} &=& -8\*s_{12} \left[
 \frac{(s_{13}+s_{23})}{\spab2.\three.1^2}
+\frac{(s_{14}+s_{24})}{\spab2.\four.1^2}
+\frac{(s_{15}+s_{25})}{\spab2.\five.1^2}
  \right] \,.
\end{eqnarray}
For the equal helicity case we have,
\begin{eqnarray}
c^{++}_{1\times3} &=& \frac{8\*\spab1.\three.1}{\spa1.2^2} \,, \\
c^{++}_{1\times23} &=& -\frac{8\*\spab1.(2+\three).1}{\spa1.2^2} \,, \\
c^{++}_{1\times2} &=& 0 \,, \\
c^{++}_{3\times12} &=& 0 \,.
\end{eqnarray}

\section{Results}
\label{sec:results}

The results of the previous two sections fully specify the amplitudes when put together with
a library for evaluating the one-loop integrals, for which we use OneLOop~\cite{vanHameren:2010cp}.
The matrix element squared
for this process, summed over colors and including the final-state symmetry factor of $1/6$,  
is given in terms of these amplitudes by,
\begin{equation}
|M^2| =  \frac{(N^2-1)}{6} \left( \frac{g_s^2}{16\pi^2} \, \frac{\mt^4}{v^3} \right)^2
 \sum_{h_1,h_2} \left| A_{\rm tot}^{h_1 h_2} \right|^2 \,.
\end{equation}
After forming this matrix element squared we cross-checked our result against OpenLoops~\cite{Buccioni:2019sur}
and Recola2~\cite{Denner:2017wsf}, finding full agreement.
Our evaluation of the matrix element is an order of magnitude or more faster than both fully-numerical codes.
Our Fortran implementation of the results in this paper is attached as ancillary files.
A Python-readable version of the coefficients of section~\ref{sec:pentagon}
can be found in the repository Antares-Results~\cite{antares_results}.

The matrix element has been implemented in the
MCFM~\cite{Campbell:1999ah,Campbell:2011bn,Campbell:2015qma,Campbell:2019dru}
parton-level event generator in order to provide cross-section predictions
for the $gg \to HHH$ process at leading order. 
For our cross-section results we use $\mh = 125$~GeV and consider the effect of both
top and bottom quarks circulating in the loop with $m_t = 173$~GeV and $m_b=4.77$~GeV.\footnote{
In the evaluation of the contribution of the bottom quark it is
appropriate to use a running mass at a scale of order $\mh$.
By instead using the pole mass we obtain an upper bound on the size
of this contribution.}
We adopt the PDF set {\tt PDF4LHC21}~\cite{PDF4LHCWorkingGroup:2022cjn}, fixing
renormalization and factorization scales to $m_{HHH}/2$.  It is useful
to express the triple- and quartic-Higgs coupling factors in terms of their
deviation from the Standard Model, $\kappa_3 = 1+ \Delta\kappa_3$,
$\kappa_4 = 1+ \Delta\kappa_4$, where $\Delta\kappa_3= \Delta\kappa_4 = 0$
in the SM.  The SM cross-sections at the 14 TeV LHC and a future
proton-proton collider operating at 100~TeV are, at this order,
\begin{eqnarray}
\sigma_{SM}^{14\text{TeV}} &=& 0.0512~\text{fb} \,, \nn \\
\sigma_{SM}^{100\text{TeV}} &=& 2.76~\text{fb} \,.
\end{eqnarray}
Note that if we had neglected the contribution of the bottom quark in the loop then
these cross sections would only be reduced by approximately $0.2$\%.
Using our code it is straightforward to compute the dependence of the cross section
on the value of the multi-Higgs couplings.
We can parametrize this dependence as,
\begin{eqnarray}
\left[ \sigma/\sigma_{SM} \right]^{14\text{TeV}} &=&
1 - 0.85 (\Delta\kappa_3) - 0.092 (\Delta\kappa_4) + 0.86 (\Delta\kappa_3)^2 \nn \\
  &-& 0.17 (\Delta\kappa_3 \Delta\kappa_4) + 0.017 (\Delta\kappa_4)^2
    -0.26  (\Delta\kappa_3)^3 \nn \\
  &+& 0.048 (\Delta\kappa_3)^2 \Delta\kappa_4 + 0.039 (\Delta\kappa_3)^4 \,,
\end{eqnarray}
and,
\begin{eqnarray}
\left[ \sigma/\sigma_{SM} \right]^{100\text{TeV}} &=&
1 - 0.66 (\Delta\kappa_3) -0.11 (\Delta\kappa_4) + 0.71 (\Delta\kappa_3)^2 \nn \\
  &-& 0.14 (\Delta\kappa_3 \Delta\kappa_4) + 0.015 (\Delta\kappa_4)^2
    -0.20 (\Delta\kappa_3)^3 \nn \\
  &+& 0.039 (\Delta\kappa_3)^2 \Delta\kappa_4 + 0.029(\Delta\kappa_3)^4 \,.
\end{eqnarray}
This parametrization agrees with the result given in Ref.~\cite{Bizon:2018syu} at 100~TeV
while the 14~TeV result differs from that in Ref.~\cite{Abouabid:2024gms} in the sign of the
$ (\Delta\kappa_4)$ term.

\begin{figure}
\begin{center}
\includegraphics[width=0.6\textwidth,angle=270]{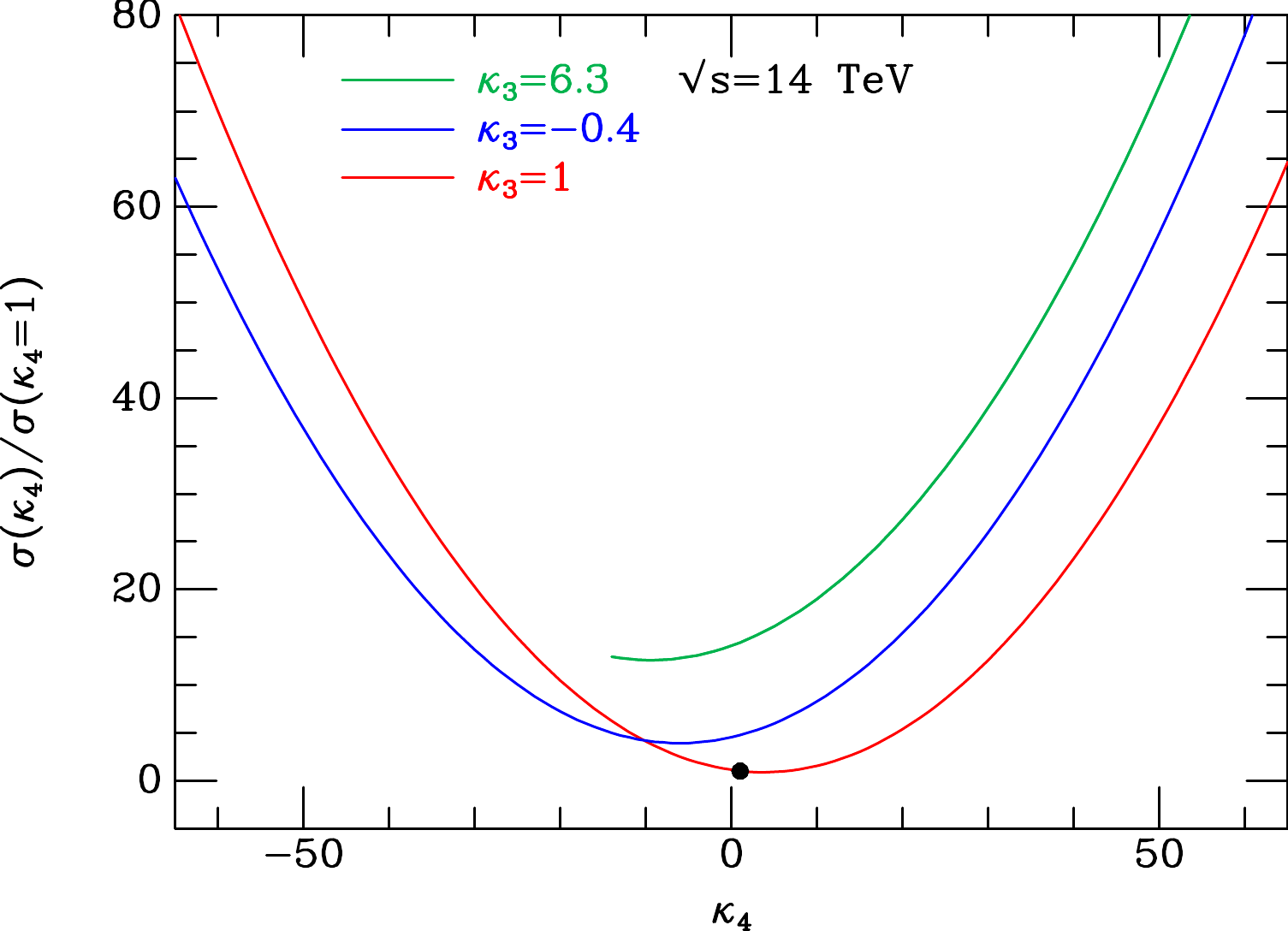}\
\caption{Dependence on $\kappa_4$ at $\sqrt{s}=14$~TeV for $\kappa_3=1$ and
the limit values of $\kappa_3$~\cite{ATLAS:2022jtk}. The curves have been restricted to the values of
$\kappa_3,\kappa_4$ allowed by perturbative unitarity~\cite{Stylianou:2023xit}.
The standard model value is shown by the black point.}
\label{kappa14}
\end{center}
\end{figure}

\begin{figure}
\begin{center}
\includegraphics[width=0.6\textwidth,angle=270]{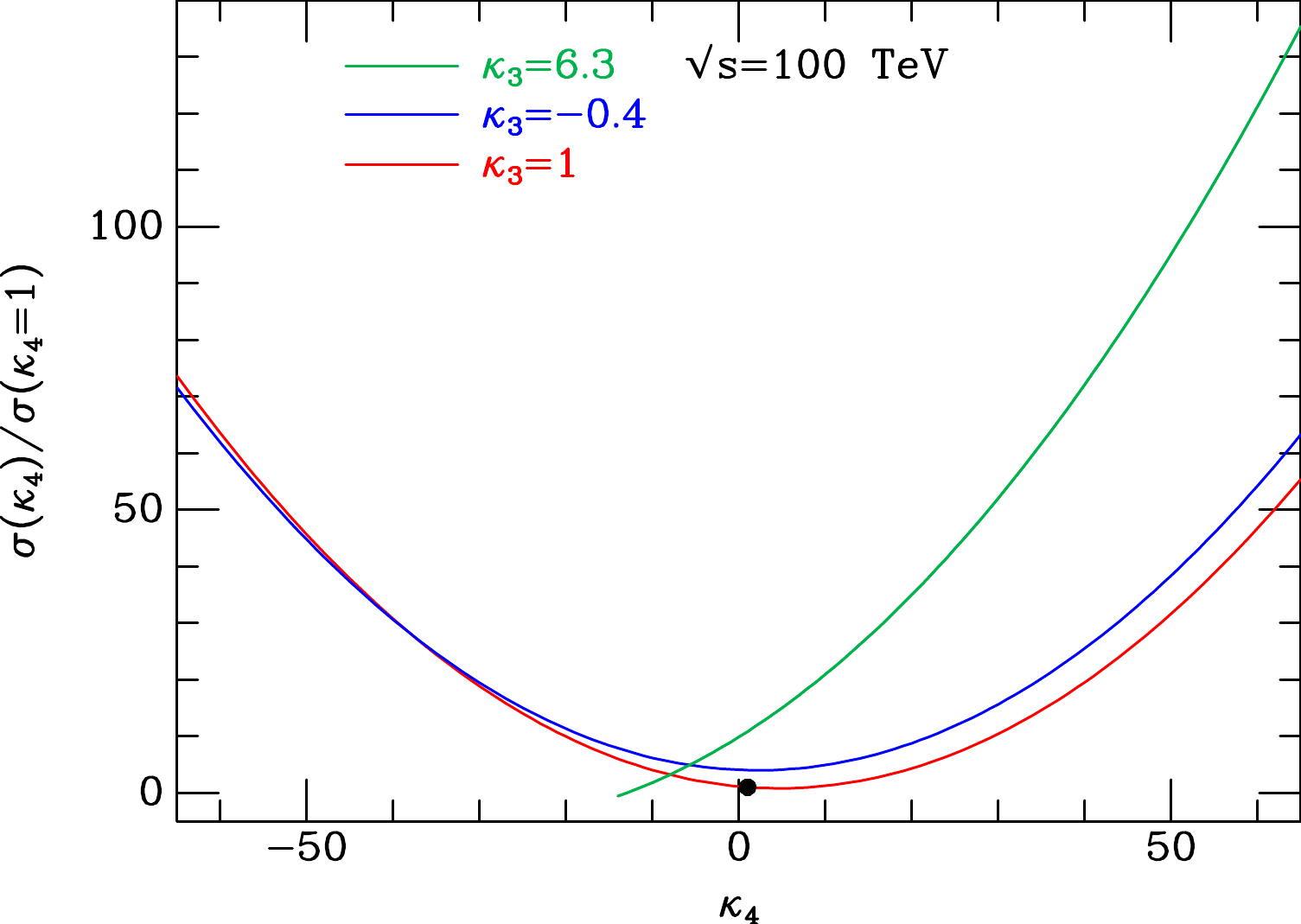}
\caption{Dependence on $\kappa_4$ at $\sqrt{s}=100$~TeV for $\kappa_3=1$ and
the limit values of $\kappa_3$~\cite{ATLAS:2022jtk}. The curves have been restricted to the values of
$\kappa_3,\kappa_4$ allowed by perturbative unitarity~\cite{Stylianou:2023xit}.
The standard model value is shown by the black point.}
\label{kappa100}
\end{center}
\end{figure}

For illustration we have plotted the dependence of the cross section on the value of $\kappa_4$ for
various choices of $\kappa_3$ in figures~\ref{kappa14}~(14~TeV) and~\ref{kappa100}~(100~TeV). 
The curves correspond to $\kappa_3=1$ and the ATLAS limit values of $\kappa_3$~\cite{ATLAS:2022jtk}
and they have been restricted to the values
of $\kappa_4$ allowed by perturbative unitarity~\cite{Stylianou:2023xit}.


\section{Conclusions}
\label{sec:conc}

We have presented a calculation of the amplitude for the process $gg \to HHH$ in full analytic form.
Despite the presence of pentagon diagrams the results can be presented in a relatively compact manner
thanks to the use of analytic reconstruction
techniques~\cite{Laurentis:2019bjh,DeLaurentis:2022otd,DeLaurentis:2023qhd}
to simplify individual integral coefficients.  Furthermore, we have massaged the coefficients
of box and pentagon integrals in order to ensure Gram determinant factors (and their square roots) cancel explicitly as
much as possible.  The resulting amplitudes are an order of magnitude faster than
equivalent numerical implementations~\cite{Denner:2017wsf,Buccioni:2019sur} and are expected to be considerably
more stable in double precision arithmetic.

Owing to its speed and stability, this amplitude is expected to be a useful ingredient in a future
NLO calculation of triple Higgs production.  An eventual 2-loop calculation of the virtual
corrections to this process could be simplified using the techniques presented here.
The methods developed in this paper also provide a blueprint for the analytic calculation of
the real radiation amplitudes, $0 \to q\bar q g HHH$ and $0 \to ggg HHH$.
Our fast amplitude calculation will also be beneficial in exploring machine-learning techniques
to discriminate between multi-Higgs production processes and their numerous
backgrounds~\cite{Abouabid:2024gms,Frank:2025zmj}.

\appendix
\section{Loop integral definitions}
\label{Integrals}
We work in the Bjorken-Drell metric so that
$l^2=l_0^2-l_1^2-l_2^2-l_3^2$.
The propagator denominators are defined as $ d_i=(l+q_i)^2-\mt^2+i\varepsilon$.
The offset momenta $q_i$
are given by sums of the external momenta, $p_i$, 
where $q_n\equiv \sum_{i=1}^n p_i$ and $q_0 = 0$.
The definition of the relevant scalar integrals is as follows,
\begin{eqnarray} \label{eq:scalarintegrals}
&& C_0(p_1,p_2;\mt)  =
\frac{1}{i \pi^{2}}
\int \, 
 \frac{d^4 l}{d_0 \; d_1 \; d_2}\, ,\nn \\
&&D_0(p_1,p_2,p_3;\mt)
= \frac{1}{i \pi^{2}}\int \, \frac{d^4 l} {d_0 \; d_1 \; d_2\; d_3}\, ,\nn \\
&&E_0(p_1,p_2,p_3,p_4;\mt)
= \frac{1}{i \pi^{2}}\int \,\frac{d^4 l}{d_0 \; d_1 \; d_2\; d_3\; d_4}\, .
\end{eqnarray}
We note that no scalar bubble integrals enter the amplitudes computed here.
For the purposes of this paper we take the masses in the
propagators to be real.  (The small imaginary part which fixes the analytic
continuations is specified by $+i\,\varepsilon$).

\acknowledgments

RKE acknowledges receipt of a Leverhulme Emeritus Fellowship from the
Leverhulme Trust. 
GDL's work is supported in part by the U.K.\ Royal Society through
Grant URF\textbackslash R1\textbackslash 20109.
The work of JMC is supported in part by the U.S. Department of Energy, Office of Science, Office of Advanced Scientific Computing Research, Scientific
Discovery through Advanced Computing (SciDAC-5) program, grant ``NeuCol''.
This manuscript has been authored by FermiForward Discovery Group, LLC under
Contract No. 89243024CSC000002 with the U.S. Department of Energy,
Office of Science, Office of High Energy Physics.

\bibliography{HHH.bib}
\bibliographystyle{JHEP}
\end{document}